\documentclass[conference]{IEEEtran}
\IEEEoverridecommandlockouts

\usepackage{cite}
\usepackage{amsmath,amssymb,amsfonts}
\usepackage{algorithmic}
\usepackage{graphicx}
\usepackage{textcomp}
\usepackage{balance}
\usepackage{xcolor}
\def\BibTeX{{\rm B\kern-.05em{\sc i\kern-.025em b}\kern-.08em
    T\kern-.1667em\lower.7ex\hbox{E}\kern-.125emX}}

    \usepackage{ragged2e}
\usepackage{floatrow}
\usepackage{tabu}
\usepackage{multirow}
\usepackage{amsthm}
\usepackage[utf8]{inputenc}
\usepackage[english]{babel}
\theoremstyle{remark}
\usepackage[export]{adjustbox}

\usepackage{multicol}

\usepackage{epstopdf}

\usepackage{float}
\usepackage{perpage}
\MakeSorted{figure}
\MakeSorted{table}
\usepackage[normalem]{ulem}
\usepackage{array}

\usepackage{graphicx}

\usepackage{amsfonts}
\usepackage{amssymb}
\usepackage{soul}
\let\labelindent\relax
\usepackage{enumitem}							
\usepackage[english]{babel}
\usepackage[utf8]{inputenc}
\usepackage[linesnumbered,ruled,vlined]{algorithm2e}
\usepackage{upgreek}
\usepackage{bm} 

\usepackage{float}
\floatstyle{plaintop}
\restylefloat{table}
\usepackage{hyperref}
\usepackage{tikz}
\usetikzlibrary{arrows.meta}
\usepackage{pifont}
\usepackage{makecell}


\usepackage{pgfplots}

\usepackage{cite}
\DeclareMathAlphabet\mathbfcal{OMS}{cmsy}{b}{n}
\ifCLASSINFOpdf
\else
\fi

\usepackage{array}
\usepackage{fixltx2e}

\usepackage[linesnumbered,ruled,vlined]{algorithm2e}

\usepackage[font=footnotesize]{caption} 
\usepackage{subcaption}
\usepackage{fix-cm}
\usepackage{comment}

\usepackage{mathtools}
\usepackage{pifont}

\usepackage{mathtools}
\usepackage{pifont}
\newcommand{\Amat}{\mathbf{A}}

\newcommand{\Hmat}{\mathbf{H}}

\newcommand{\Rmat}{\mathbf{R}}

\newcommand{\av}{\mathbf{a}}

\newcommand{\cv}{\mathbf{c}}

\newcommand{\gv}{\mathbf{g}}

\newcommand{\pv}{\mathbf{p}}
\newcommand{\qv}{\mathbf{q}}

\newcommand{\wv}{\mathbf{w}}
\newcommand{\xv}{\mathbf{x}}
\newcommand{\yv}{\mathbf{y}}
\newcommand{\zv}{\mathbf{z}}

\newcommand{\alphav}{\boldsymbol{\alpha}}
\newcommand{\phiv}{\boldsymbol{\phi}}

\newcommand{\etav}{\boldsymbol{\eta}}

\newcommand{\gammav}{\boldsymbol{\gamma}}

\usepackage{tikz}

\begin{document}
\raggedbottom
\bstctlcite{IEEEexample:BSTcontrol}


\title{ Phase-Coherent D-MIMO ISAC: Multi-Target Estimation and Spectral Efficiency Trade-Offs}


\author{\\[-20pt]\IEEEauthorblockN{ {Venkatesh Tentu, Henk Wymeersch, Musa Furkan Keskin, Sauradeep Dey, Tommy Svensson } }
\IEEEauthorblockA{ Department of Electrical Engineering, Chalmers University of Technology, Sweden
\\ \{tentu, henkw, furkan, deysa, tommy.svensson\}@chalmers.se
}
\thanks{This work was supported by the SNS JU projects Hexa-X-II (Grant Agreement No. 101095759), 6G-DISAC (Grant Agreement No. 101139130) under the EU’s Horizon Europe Research and Innovation Programme,
the Advanced Digitalization program at the WiTECH Centre DisCouRSe financed by VINNOVA, Chalmers, Ericsson, Qamcom, RISE, SAAB and Volvo Cars,
and the Swedish Research Council (VR) through
the project 6G-PERCEF under Grant 2024-04390. The computations were enabled by resources provided by the National Academic Infrastructure for Supercomputing in Sweden (NAISS). } 
\vspace{-0.9cm}
}

\bstctlcite{IEEEexample:BSTcontrol}
\maketitle

\begin{abstract}
We investigate distributed multiple-input multiple-output (D-MIMO) integrated sensing and communication (ISAC) systems, in which multiple phase-synchronized access points (APs) jointly serve user equipments (UEs) while cooperatively detecting and estimating multiple static targets. To achieve high-accuracy multi-target estimation, we propose a two-stage sensing framework combining non-coherent and coherent maximum-likelihood (ML) estimation.
In parallel, adaptive AP mode-selection strategies are introduced to balance communication and sensing performance: a communication-centric scheme that maximizes downlink spectral efficiency (SE) and a sensing-centric scheme that selects geometrically diverse receive APs to enhance sensing coverage. 
Simulation results confirm the SE–sensing trade-off, where appropriate power allocation between communication and sensing, and larger array apertures alleviate performance degradation, achieving high SE with millimeter-level sensing precision. We further demonstrate that the proposed AP-selection strategy reveals a sweet-spot number of receive APs that maximizes sensing coverage without significantly sacrificing SE.






\end{abstract}


\vspace{-3mm}
\section{Introduction}
Integrated sensing and communication (ISAC) is a key enabler for sixth-generation (6G) networks, where the radio infrastructure jointly provides connectivity and environmental awareness~\cite{Fan_LIU_2023_Magazine,Zhang_2022_ISAC_Survey}. 
Conventional monostatic and bistatic ISAC architectures efficiently reuse spectrum but remain limited by array aperture size, angular diversity, and blockage. 
Distributed multiple-input multiple-output (D-MIMO) architectures overcome these constraints by employing spatially separated access points (APs) that jointly illuminate and observe the environment~\cite{Henk_Radio_Stripes_Journal_25,demirhan_24_CF_ISAC, Elfiatoure_2025-CF-ISAC}. By exploiting network-wide spatial diversity, D-MIMO ISAC can substantially improve sensing accuracy while maintaining communication reliability~\cite{Elfiatoure_2025-CF-ISAC,Christos_UCL_2025_DMIMO}.


Recent works have investigated the potential of D-MIMO ISAC from various perspectives. On the transmitter side, prior studies optimized resource allocation and AP mode selection. 
For instance, adaptive AP mode selection and power allocation were employed to balance downlink spectral efficiency (SE) and sensing-zone detection~\cite{Elfiatoure_2025-CF-ISAC}, while a graph-learning-based cell-free ISAC framework \cite{Jiang_Peng_2025_ML_AP_selection} optimized AP mode selection without explicitly analyzing the impact of the transmit–receive AP configuration on ISAC trade-offs.
A joint power-minimization under UE signal-to-interference-plus-noise-ratio (SINR) and sensing Cramér–Rao lower bound (CRLB) constraints was presented in~\cite{Huang_Yi_2022_DMIMO}, while~\cite{demirhan_24_CF_ISAC} proposed a joint sensing and communication beamforming design for D-MIMO ISAC systems. In addition, several works  studied cooperative networks that utilize dedicated sensing and communication signals to perform target detection and serve communication UEs~\cite{Zinat_24_Multi_Static_DMIMO,Gaoyuan_Cheng_2024_DMIMO}, and user-centric association frameworks were extended to distributed ISAC scenarios to ensure sufficient sensing geometry around each UE~\cite{Naeem_2025_UE_association_ISAC}. Non-coherent ISAC designs were further analyzed in~\cite{NON_COH_CHRISTOS}, evaluating the sensing–communication trade-off via CRLB and SINR metrics.
Although these studies (except~\cite{Jiang_Peng_2025_ML_AP_selection}) capture the fundamental coupling between communication and sensing, they primarily evaluate sensing performance through performance bounds or detection metrics, considering phase coherence only from the transmit side.


On the receiver side, several works have investigated signal processing and estimation methods for distributed and multistatic ISAC systems, where distributed APs jointly process received echoes for detection or localization~\cite{Jiang_Peng_2025_ML_AP_selection, Bauhofer_2023_6Gnet,Shiraki_2023_Localization,Elisabetta_Matricardi_2025_DMIMO_OFDM}. 
OFDM-based multistatic architectures with time-synchronized receivers were experimentally validated in~\cite{Bauhofer_2023_6Gnet}, while \cite{Shiraki_2023_Localization} analyzed single-carrier narrowband systems.  In \cite{Elisabetta_Matricardi_2025_DMIMO_OFDM}, distributed APs jointly performed multistatic MIMO-OFDM sensing for multi-target localization.
These works demonstrate the potential of cooperative receiver processing under time-synchronization; however, their phase-asynchronous operation leads to non-coherent fusion and confines them to sensing-only functionality without communication integration.
Although \cite{Jiang_Peng_2025_ML_AP_selection} integrated multi-target estimation and communication via a data-driven framework, it relied on implicit neural inference, without explicit phase-coherent modeling or analytical performance characterization.
While non-phase-synchronous operation simplifies implementation, it prevents coherent echo fusion. In contrast, phase-synchronized processing with a common AP phase reference enables inter-AP phase exploitation for improved sensing accuracy, yet remains unaddressed in distributed ISAC frameworks.

In summary, existing D-MIMO ISAC works either optimize system-level communication and sensing trade-offs without explicit target estimation~\cite{Elfiatoure_2025-CF-ISAC, Jiang_Peng_2025_ML_AP_selection, Huang_Yi_2022_DMIMO, demirhan_24_CF_ISAC, Zinat_24_Multi_Static_DMIMO, Gaoyuan_Cheng_2024_DMIMO, Naeem_2025_UE_association_ISAC, NON_COH_CHRISTOS} or focus on target estimation using time-synchronized APs without integrating communication~\cite{ Bauhofer_2023_6Gnet,Shiraki_2023_Localization,Elisabetta_Matricardi_2025_DMIMO_OFDM}.
Moreover, prior multi-static D-MIMO ISAC works do not exploit receiver-side phase coherence for target estimation, unlike D-MIMO radar studies that typically consider coherent receivers.
 Existing AP mode selection strategies {did not} explore the {explicit}  communication and sensing trade-offs~\cite{Elfiatoure_2025-CF-ISAC,Jiang_Peng_2025_ML_AP_selection}.  
To address these gaps, we propose a phase-coherent D-MIMO ISAC framework where  synchronized half-duplex APs jointly serve UEs and cooperatively detect and estimate multiple passive targets with unknown positions. We develop a two-stage maximum-likelihood (ML) estimator that first performs non-coherent coarse detection, followed by coherent refinement for wavelength-level accuracy. 
Under half-duplex constraint, we further introduce AP-mode selection strategies, where each AP dynamically switches between transmit and receive modes to  balance communication and sensing.
We introduce 
two greedy AP-mode selection strategies (a communication-centric and a sensing-centric approach), and demonstrate that sacrificing a small number of APs for sensing does not impact the communication metrics but significantly improves sensing~coverage.  


\section{System model} \label{sec:system_model}
\begin{figure}
    \centering
    \includegraphics[width=0.8\linewidth]{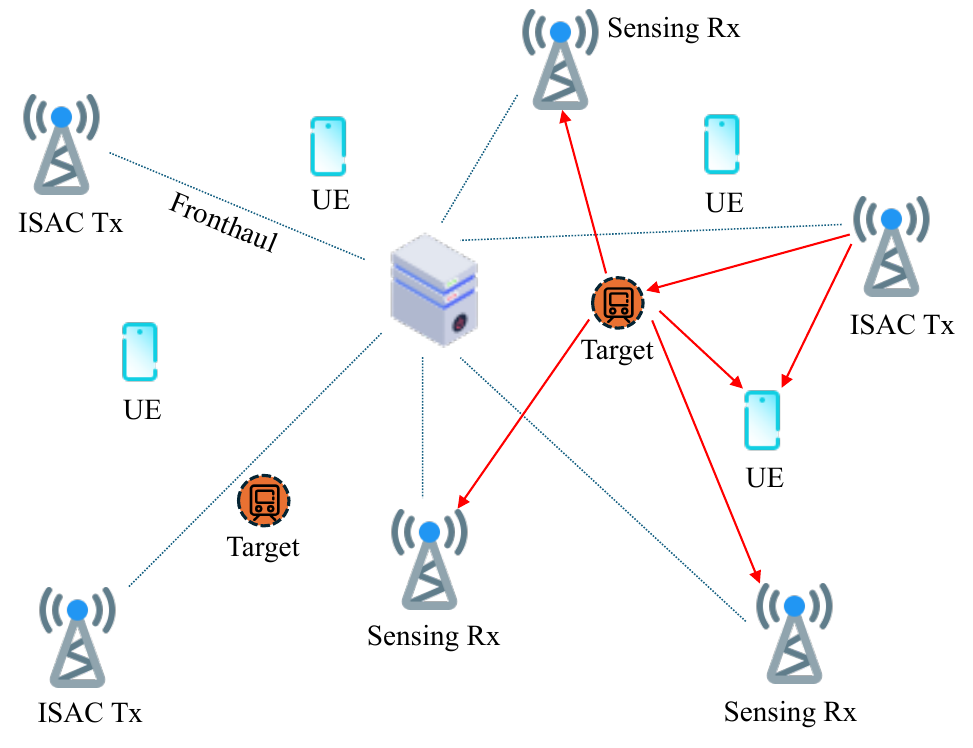} \vspace{-1mm}
    \caption{System model of a downlink D-MIMO ISAC with phase-synchronized APs for joint communication and target estimation.}
    \label{fig:SystemModel_ISAC}
\end{figure}
We consider a downlink D-MIMO ISAC system as shown in Fig.~\ref{fig:SystemModel_ISAC}, where $M$ mode-switching APs, each equipped with  a uniform linear array (ULA) of $N$ antennas, jointly communicate with $K$ single-antenna UEs and perform sensing to estimate the locations of $S$ targets. Among these, a subset $\mathcal{T}$ of $M_t$ APs is configured as transmitters, while the remaining $\mathcal{R}$ of size $M_r = M - M_t$ operate as receivers for sensing. 
We allocate separate beams for serving communication UEs and for sensing to detect multiple targets. Let $q_k[l] \in \mathbb{C}$ and $\qv_0[l] \in \mathbb{C}^{N \times  1}$   denote, respectively, the communication symbol intended for the $k$th UE and the multi-antenna sensing signal at time instant $l$.  Both symbols are assumed to be zero mean and unit power, i.e., $\mathbb{E}\{|q_k[l]|^2\}  = 1$, and $\mathbb{E}\{\|\qv_0[l]\|^2\} =1$. 
Accordingly, at time instant $l$, the signal transmitted by the $t$th AP can be written as 
 \begin{align}
\xv_{t}[l]    & = \sqrt{\rho_t }  \sum_{k=1}^{K}  \sqrt{p_{tk}} \wv_{tk}  q_k[l] + \delta_t[l] \sqrt{1-\rho_t} \sqrt{P_t}      {\qv_0[l]}. \label{eq_TX signal_m} 
\end{align}
Here, $\rho_t \! \in \!  [0,1]$ controls the fraction of the $t$th AP total power allocated for the communication, and $p_{tk}$ denotes the per-UE power control coefficient. The $t$th AP transmit signal satisfies the  per-AP power constraint $P_t$, i.e., $\mathbb{E}\{\|\xv_t[l]\|^2\} \!= \rho_t \sum_{k=1}^{K} p_{tk} + (1-\rho_t)\delta_t[l] P_t \leq P_t$. The~vector~$\wv_{tk}  $ denotes the normalized maximal-ratio beamforming vector of the $t$th AP towards UE $k$, such that $\| \wv_{tk} \| =1 $.
The term $\delta_t[l] \in \{0, 1 \}$ indicates whether the $t$th AP transmits the sensing waveform at time $l$. Enforcing $\sum_{t=1}^{M_t} \delta_t[l] =1 \, \forall l$ ensures orthogonal sensing, where only one AP is active for sensing at a given instant. This eliminates inter-AP interference and provides an unambiguous transmitter association, enabling per-transmitter multistatic processing and coherent multi-AP sensing.

\subsection{Channel Model}
\subsubsection{Sensing Channel Model} 
Let $\Hmat_{t,r}^s \in \mathbb{C}^{N \times N}$ denote the sensing channel between the $t$th transmit AP and the $r$th receive AP through the target $s$, expressed as \cite{Zinat_24_Multi_Static_DMIMO} 
\begin{align} \label{eq_Sensing_channel}
    \Hmat_{t,r}^s &= \alpha_{t,r}^s {e^{j \varphi_{t,r}^s}} \av\big( \vartheta_r^s \big) \av^{T}\big( \theta_t^s\big). 
\end{align}
The term $\alpha_{t,r}^s \in \mathbb{R}$ represents the combined sensing channel gain, including the loss in the two-leg path from the $t$th transmit AP to the $s$th target to the $r$th receive AP and the variance of the radar cross section (RCS) of the $s$th target $\sigma_{\text{rcs}}^s$. 
Since all APs are phase-synchronized, the oscillator-induced phase offsets are compensated.
Accordingly, the carrier phase term $\varphi_{t,r}^s$ accounts only for the  propagation delay and scattering-induced phase shift from the $s$th target, and is given as 
    $\varphi_{t,r}^s = -2\pi f \tau_{t,r}^s +  \phi_{t,r}^s$,   
%
where $\tau_{t,r}^s = ({\| \pv^s - \pv^{\text{AP}}_{t} \| + \|  \pv^{\text{AP}}_{r} - \pv^s \|})/{c}$ accounts for the propagation delay of the two-leg path, with $\pv^s$, $\pv^{\text{AP}}_{t} $ and $\pv^{\text{AP}}_{r} $ being the 2D position of the $s$th target, and $t$th transmit and $r$th receive AP.  
Under isotropic reflectivity, the scattering-induced phase shift $\phi_{t,r}^s $ is assumed constant across all AP pairs, i.e.,  $\phi_{t,r}^s = \phi^{\text{fix},s} $.
The terms $\av\big( \theta_t^s\big)$ and $\av\big( \vartheta_r^s\big)$ in \eqref{eq_Sensing_channel} denote the transmit and receiver array response vectors of the $t$th transmit AP and the $r$th receive AP to the  target $s$. The vector $\av\big( \theta\big)$, with $\theta \in \{\theta_t^s, \vartheta_r^s \}$, can be given as
\begin{align}
  \av\big( \theta\big) & = \big [ 1, e^{j \frac{2 \pi d}{\lambda}  \text{sin}(\theta)} , \cdots,  e^{ j (N-1) \frac{2 \pi d}{\lambda}  \text{sin}(\theta)} \big]^\top.
\end{align}
The angles of departure (AoD) $\theta_{t}^s$ and angles of arrival (AoA) $\vartheta_{r}^s$ are measured relative to each AP's boresight orientation, and determined by the AP-target geometry: $\theta_{t}^s = ( \bar{\theta}_{t}^s - \theta_t^{\text{AP}} ) \text{mod}(2 \pi)$, and $\vartheta_{r}^s =  ( \bar{\vartheta}_{r}^s - \vartheta_r^{\text{AP}})\text{mod}(2 \pi)$.
Here, $\bar{\theta}_{t}^s \!=  \text{atan2}\Big( \frac{[ \pv^{\text{AP}}_{t} - \pv^s ]_y}{[ \pv^{\text{AP}}_{t} - \pv^s ]_x} \Big)$ and similar definitions apply~for~$\bar{\vartheta}_{r}^s$.

\subsubsection{Communication Channel Model} 
The downlink communication channel $\gv_{tk}$ between the $t$th transmit AP to the $k$th UE is modeled as~\cite{Venkatesh_CF_2022_Channel}
\begin{align}
\mathbf{g}_{tk}=   \mathbf{\bar{g}}_{tk} +  \tilde{\mathbf{g}}_{tk}, \label{channel model}
\end{align}
where $\mathbf{\bar{g}}_{tk}$ and $\tilde{\mathbf{g}}_{tk}$ represent the deterministic and stochastic non-line-of-sight (sNLoS) components, respectively.
The sNLoS component is distributed as  $\tilde{\mathbf{g}}_{tk} \sim \mathcal{C}\mathcal{N}\left(0, \Rmat_{tk} \right)$, with $\Rmat_{tk} \in \mathbb{C}^{N \times N}$ being the spatial-correlation matrix. The deterministic component $\mathbf{\bar{g}}_{tk}$ is expressed as
\begin{align}
    \mathbf{\bar{g}}_{tk}   =  \bar{\beta}_{tk} {e^{-j2\pi f \tau_{tk}} e^{ j{\phi_{k} }  } } \av\big( \theta_t^k\big) 
    + \sum_{s=1}^S \varpi_{t,k}^s {e^{j \varsigma_{t,k}^s}}   \av\big( \theta_t^s\big). \label{channel model_LoS_Comm}
\end{align}
The first term is the direct LoS path, and the second term captures the reflected channel component corresponding to the indirect AP–target–UE propagation path, similar to the sensing channel described in~\eqref{eq_Sensing_channel}.
 Here, $\bar{\beta}_{tk}$ and $\tau_{tk} $ are the LoS channel gain and path delay between the $t$th AP and 
 the $k$th UE, respectively.  
 In addition, $\phi_{k} \sim \mathcal{U}(-\pi, \pi)$ models the unknown constant phase offset between the $k$th UE's local oscillator (LO) and the phase-coherent transmit APs' LO reference.
Similar to the sensing case, $\varpi_{t,k}^s$ denotes the combined channel gain, $\varsigma_{t,k}^s$ is the carrier phase, and  the vector $\av\big( \theta_t^k\big)$ represents the transmit array response, with AoD $\theta_{t}^k$. 


\vspace{-2pt}
\subsection{Communication SE}\label{sec:commSE}
\vspace{-2pt} The signal received at the $k$th UE during time instant $l$, using \eqref{eq_TX signal_m}, is $y_k[l]  = \sum_{t=1}^{M_t} \gv_{tk}^{\mathsf{H}} \xv_t[l] + z_k[l] $. We expand it as
\begin{align}
     y_k[l]   & =   \sum_{t=1}^{M_t} \gv_{tk}^{\mathsf{H}} \big( \sqrt{\rho_t p_{tk}} \wv_{tk}   \big)  q_k[l] +    \sum_{t=1}^{M_t}  \delta_t[l] \! \sqrt{P_t(\rho)}    \gv_{tk}^{\mathsf{H}}  \qv_0[l] \notag \\ \vspace{-3pt}
    &  \quad  +  \sum_{i \neq k}^{K} \sum_{t=1}^{M_t} \gv_{tk}^{\mathsf{H}} \big( \sqrt{\rho_t p_{tk}} \wv_{ti}) q_i[l]    +  z_k[l]. \label{Eq_UE_Rx_k}
\end{align}
Here, the first, second, and third terms denote the desired signal,  the interference from the sensing signal at the $k$th UE, and the multi-user interference, respectively. The terms $P_t(\rho) = (1-\rho_t)P_t$, and $z_k[l] \sim \mathcal{CN}(0, \sigma_k^2)$ is the additive white Gaussian noise (AWGN) at the UE $k$.  
Using \eqref{Eq_UE_Rx_k}, the achievable downlink SE for $k$th UE is given as $\text{SE}_k = \text{log}_2 (1+ \text{SINR}_k )$, where~$\text{SINR}_k  =$
\begin{align}
  \! \!   \frac{ \Big|     \sum\limits_{t=1}^{M_t} \sqrt{\rho_t p_{tk}}  \gv_{tk}^{\mathsf{H}}    \wv_{tk}      \Big|^2 }{ \sum\limits_{i \neq 1}^{K}  \! \Big|  \sum\limits_{t=1}^{M_t} \!\sqrt{\! \rho_t p_{tk}}  \gv_{tk}^{\mathsf{H}}    \wv_{ti}  \Big|^2 \!+  \! \frac{1}{N} \Big\| \sum\limits_{t=1}^{M_t} \!  \delta_t[l] \! \sqrt{P_t(\rho)}    \gv_{tk} \Big\|^2 \!+ \sigma_k^2 }.\!
\end{align}
The above expression is derived by assuming independence among the transmit data symbols, sensing signals, and noise. 

\section{Joint Target Estimation} 
In this section, we estimate the 2D target positions by exploiting the multiple transmissions over time. The $t$th transmit AP sends $\xv_t[l]$ in \eqref{eq_TX signal_m} at the $l$th time instant. The echo signal reflected from the $S$  targets received at the AP $r$ is given by  
\begin{align} \label{eq_sensing_Rx}
 \! \! \! \yv_{r}[l]  = \sum_{s=1}^{S} \sum_{t =1}^{M_t} \Big( \alpha_{t,r}^s {e^{j \varphi_{t,r}^s}}  \av\big( \vartheta_r^s \big) \av^{T}\big( \theta_t^s\big) \Big)  \xv_{t}[l]   + \zv_r[l].   
\end{align}
Here, $\zv_r[l]$ is AWGN at the $r$th AP. Each receiver AP sends its respective signal $\yv_{r}[l]$ to a central processing unit (CPU) for joint processing. Given the observations $\{ \yv_{r}[l] \}_{r=1}^{M_{r}}$, the goal is to estimate the target positions $\pv_{\text{2D}}$. The unknown parameter vector is $\etav = [\pv_{\text{2D}}^\top, \alphav^\top, (\phiv^{\text{fix}})^\top ]^\top \in \mathbb{R}^{3S + M_tM_rS}$, 
where $\pv_{\text{2D}} \!=\! [p_{11}, p_{12}, \cdots, p_{S1}, p_{S2}]^\top \! \in \mathbb{R}^{2S \times 1}$,  $\alphav \! = [\alpha_{1,1}^1, \cdots, \alpha_{M_t,M_r}^S ]^\top \! \in \mathbb{R}^{M_tM_rS \times 1}$ and $\phiv^{\text{fix}} \!=  [\phi^{\text{fix},1}, \cdots, \phi^{\text{fix},S}]^\top \in \mathbb{R}^{S \times 1}$. 
For a given $\etav$, the target estimation problem is formulated as an ML log-likelihood estimation:
 \begin{align}
     \hat{\etav} = \text{arg } \underset{\etav \in \mathbb{R}^{3S + M_tM_rS}}{\text{max}} \text{ log} \; p \big(   \yv^{(s)}| \etav   \big). \label{eq_PDF_y_sensing}
 \end{align}
Solving \eqref{eq_PDF_y_sensing} is challenging due to its highly non-linear and non-convex structure. The dependence of 
carrier phase terms on the unknown target positions results in multiple local optima, making it challenging to obtain a reliable closed-form or globally optimal solution. To ensure high sensing accuracy, 
a low-complexity two-stage estimator framework is adopted, combining non-coherent and coherent processing.\vspace{-1pt}
\subsection{Coherent ML Estimation } 
We formulate the multi-target position estimation problem as a joint estimation over all targets, i.e., estimating $\pv_{\text{2D}}$ by exploiting the dependence of the carrier phase $\varphi_{t,r}^s$ on the target positions. Neglecting constant terms, \eqref{eq_PDF_y_sensing} becomes:
 \begin{align}
     \hat{\etav} = \arg \underset{\etav \in \mathbb{R}^{3S + M_tM_rS}}{\min}  \mathcal{L}(\etav ), \label{eq_PDF_y_sensing_mean} 
      \end{align}
      \vspace{0.1em} 
      where 
       \begin{align}
      \mathcal{L}(\etav) \! = \! \sum_{l=1}^{L} \sum_{r=1}^{M_r} \!  \Big\|    \yv_r[l] \!-  \! \sum_{t=1}^{M_t} \sum_{s=1}^{S} \alpha_{t,r}^s {e^{j \varphi_{t,r}^s}}  \av\big( \vartheta_r^s \big) \av^{T}\big( \theta_t^s\big)   \xv_{t}[l]  \Big\|^2_2\!. \notag
 \end{align}
We first solve the above problem to find the optimal values of $\alpha_{t,r}^s \forall t,r,s$ and scattering-induced phase shift $\phi^{\text{fix},s}$.
The parameter $\alpha_{t,r}^s$ can be estimated by differentiating the ML objective function in \eqref{eq_PDF_y_sensing_mean} and equating it to zero as:
\begin{align}
    \widehat{\alpha}_{t,r}^s = \frac{ \sum_{l=1}^{L} \mathcal{R} \Big( (\yv_r[l])^{\mathsf{H}}  {e^{j \varphi_{t,r}^s}}  \av\big( \vartheta_r^s \big) \av^{T}\big( \theta_t^s\big) \xv_{t}[l] \Big) }{   \sum_{l=1}^{L} \| \av\big( \vartheta_r^s \big) \av^{T}\big( \theta_t^s\big)\xv_{t}[l] \|^2 }. \label{eq_Alpha_estimate}
\end{align}
Before obtaining the ML estimate for the scattering-induced phase shift $\phi^{\text{fix},s}$, we re-express the ML objective function in \eqref{eq_PDF_y_sensing_mean} by substituting $\hat{\alpha}_{t,r}^s$ from \eqref{eq_Alpha_estimate} as $\mathcal{L}( \pv_{\text{2D}}, \phiv^{\text{fix}} )$ given in \eqref{eq_CP_ML_after_Alpha_estimation} at the top of the page.
Here $\Amat( \vartheta_r^s, \theta_t^s ) =  \av\big( \vartheta_r^s \big) \av^{T}\big( \theta_t^s\big)$.
\begin{figure*}[t!]
\setcounter{equation}{11} 
\vspace{3pt}
\footnotesize
\begin{align} 
\!\! \mathcal{L}( \pv_{\text{2D}}, \phiv^{\text{fix}} ) =    \sum_{l,r}  \Bigg\|   \Bigg(  \underbrace{\yv_r[l] - \sum_{s,t}     \frac{ \sum\limits_{l'} \big[ \Amat( \vartheta_r^s, \theta_t^s)    \xv_{t}[l'] \big]^{\mathsf{H}} \yv^{H}_r[l']  }{ 2\sum\limits_{l'} \| \av\big( \vartheta_r^s \big) \av^{T}\big( \theta_t^s\big)\xv_{t}[l'] \|^2 } \Amat( \vartheta_r^s, \theta_t^s)     \xv_{t}[l] }_{ \check{\yv}_r[l] }
- \sum_{s,t}  {e^{j 2\varphi_{t,r}^s}}  \underbrace{ \frac{    \sum\limits_{l'}   \yv^{H}_r[l']  \Amat( \vartheta_r^s, \theta_t^s) \xv_{t}[l'] }{   2\sum\limits_{l'} \| \av\big( \vartheta_r^s \big) \av^{T}\big( \theta_t^s\big)\xv_{t}[l'] \|^2 }    \Amat( \vartheta_r^s, \theta_t^s)     \xv_{t}[l] }_{\cv_{t,r}^s[l]} \Bigg) \Bigg\|^2_2.\! \label{eq_CP_ML_after_Alpha_estimation}
\end{align}
\normalsize 
\vspace{-15pt}
\end{figure*}
We re-express \eqref{eq_CP_ML_after_Alpha_estimation} in compact form as
\begin{align}  \setcounter{equation}{12}  
\!\!\!\!\!\mathcal{L}( \pv_{\text{2D}}, \phiv^{\text{fix}} ) = \!\sum_{l=1}^{L} \sum_{r=1}^{M_r} \! \Big\|     \check{\yv}_r[l] -\! \sum_{s=1}^{S} \sum_{t=1}^{M_t}  {e^{j2 \varphi_{t,r}^s}} \cv_{t,r}^s[l]     \Big\|^2_2.\!\! \label{eq_CP_ML_2D_position}
\end{align}
From \eqref{eq_CP_ML_2D_position}, the fixed reflection phase $\widehat{\phi}^{\text{fix},s} $ can be estimated~as
\begin{align}
    \widehat{\phi}^{\text{fix},s} = \frac{1}{2} \angle \Big( \sum\limits_{l,r,t}  \check{\yv}^{H}_r[l]   {\cv_{t,r}^s[l] }  {e^{j4  \pi f \tau_{t,r}^s }}  \Big)  + A\pi, \label{eq_phase_RCS_esti} 
\end{align}
where $A \in \mathbb{Z}$ accounts for the integer ambiguities in the phase estimation. Substituting \eqref{eq_phase_RCS_esti} in \eqref{eq_CP_ML_2D_position}, we have $ \mathcal{L}(\pv_{\text{2D}} )$ as given in \eqref{eq_Objective_P2D-CP} at the top of the page.
\begin{figure*}[t!]
\setcounter{equation}{14}  
\vspace{3pt}
\footnotesize
\begin{align} 
  \mathcal{L}(\pv_{\text{2D}} ) 
  = \sum\limits_{l=1}^{L} \sum\limits_{r=1}^{M_r}  \Big(  \| \check{\yv}^{(s)}_r[l]\|^2    +  \Big\|\sum_{s=1}^{S} \sum_{t=1}^{M_t}  {e^{-j4\pi f \tau_{t,r}^s  }}   {\cv_{t,r}^s[l] }  \Big\|^2 \Big)   - 2  \Big|  \sum\limits_{l=1}^{L} \sum\limits_{r=1}^{M_r}  \sum_{s=1}^{S} \sum_{t=1}^{M_t} {e^{j2 (-2\pi f \tau_{t,r}^s ) }}  (\check{\yv}^{(s)}_r[l])^{\mathsf{H}}  {\cv_{t,r}^s[l] }   \Big|.  \label{eq_Objective_P2D-CP}
\end{align}
\normalsize
\hrule
\vspace{-15pt}
\end{figure*}
The expression for the ML position estimate becomes
\begin{align} \setcounter{equation}{15} 
 \widehat{\pv}_{\text{2D}}  = \text{arg } \; \underset{ \pv_{\text{2D}} \in \mathbb{R}^{2S \times 1}}{\text{min }} \mathcal{L}(\pv_{\text{2D}} ). \label{eq-CP_ML_2d_7}
\end{align}
Solving \eqref{eq-CP_ML_2d_7} is challenging due to its highly non-linear structure and multiple local minima. To enable efficient refinement, we employ a low-complexity non-coherent initializer that provides reliable coarse estimates, which subsequently facilitates efficient refinement of \eqref{eq-CP_ML_2d_7} via gradient-based methods.
\subsection{Non-coherent ML Estimation} 
 In this section, we develop an estimator that does not exploit the carrier phase information $\varphi_{t,r}^s$ and avoids the high-dimensional optimization and spiky cost function. 
To this end, we treat the channel amplitudes as complex variables $\gamma_{t,r}^s = \alpha_{t,r}^s e^{j \varphi_{t,r}^s}$ and formulate the single-target ML cost function without carrier phase information as 
 \begin{align}
 \!\!  \!\!  \hat{\etav}  
    = \! \underset{\bar{\etav}}{\text{argmin }} \!\!\sum_{l=1}^{L} \sum_{r=1}^{M_r} \! \Big\|   \yv_r[l] - \! \sum_{t=1}^{M_t}  \gamma_{t,r}^s \av\big( \vartheta_r^s \big) \av^{T}\!\big( \theta_t^s\big)   \xv_{t}[l]  \Big\|^2_2. \!\!\label{eq_PDF_y_sensing_mean_NCP}
 \end{align}
 The position domain parameter vector becomes $\bar{\etav} \!=\! [\bar{\pv}_{\text{2D}}^\top, \Re\{\gammav \}, \Im\{\gammav \} ] \! \in \! \mathbb{R}^{2 + 2M_tM_r}$.
Similar to~\eqref{eq_Alpha_estimate}, the estimates of complex channel gain can be obtained~as~$\smash{\widehat{\gamma}_{t,r}^{s}}  =$ \newline
    ${ \sum_{l=1}^{L} \!  \big(   \av\big( \vartheta_r^s \big) \av^{T}\big( \theta_t^s\big)    \xv_{t}[l] \big)^{\mathsf{H}} \yv_r[l]   }/{ \sum_{l=1}^{L} \! \big \| \av\big( \vartheta_r^s \big) \av^{T}\big( \theta_t^s\big)    \xv_{t}[l]   \big \|^2 }\!  $, 
leading to the compressed cost function: \vspace{0.1pt}
\begin{align}
 \mathcal{L}_{\text{NCP}}(\bar{\pv}_{\text{2D}} ) \!= \!  \sum_{l=1}^{L} \sum_{r=1}^{M_r}  \Big\|   \yv_r[l] \! - \!   \sum_{t=1}^{M_t} { \widehat{\gamma}_{t,r}^{s}} \av\big( \vartheta_r^s \big) \av^{T}\big( \theta_t^s\big)   \xv_{t}[l]  \Big\|^2_2.  \notag 
\end{align}
The final expression for the non-coherent ML estimator that does not exploit the carrier phase is
\begin{align}
  \widehat{\pv}_{\text{2D}}^{\text{NCP}}  = \text{arg } \; \underset{ \bar{\pv}_{\text{2D}} \in \mathbb{R}^2}{\text{min }} \mathcal{L}_{\text{NCP}}(\bar{\pv}_{\text{2D}} ). \label{eq-NCP_ml_2D}
\end{align}
Similar to \eqref{eq-CP_ML_2d_7}, we cannot obtain a closed-form solution. We solve this problem via a 2D grid search, evaluating the single-target non-coherent objective in \eqref{eq-NCP_ml_2D} over the entire $(x,y)$ area to form a cost map. When the targets are sufficiently separated, this cost map exhibits one dominant dip (negative peak) per actual target. We then detect these dips using a constant false alarm rate (CFAR) test and use their coordinates as coarse position hypotheses to initialize the coherent estimator in \eqref{eq-CP_ML_2d_7}.

\section{AP Mode Selection} \label{sec:APselection}
In a D-MIMO ISAC system, the selection of transmit and receive APs plays a critical role in balancing the dual objectives of communication performance and sensing coverage.  Since an exhaustive search over all possible assignments is computationally intractable, we present two heuristic AP selection strategies: communication- and sensing-centric. For both strategies, we fix the value of $M_t$ and $M_r$, with $M=M_t+M_r$.
\subsection{Communication-centric Approach}
 In the communication-centric AP selection, we aim to maximize downlink SE performance by aligning transmit APs with the strongest AP–UE channel conditions. We build a transmit AP set $\mathcal{T}$ iteratively. The selection starts from an empty set, choosing first the AP that provides the highest single-AP SE. At each iteration $i$, we evaluate the impact of adding a candidate AP on the aggregate communication SE and select the one offering the highest SE gain. We define the aggregated link strength to UE $k$ after selecting set $\mathcal{T}$ as $\beta_{\text{sum},k}(\mathcal{T}) \triangleq \sum_{m\in\mathcal{T}} \beta_{m,k}$. Here, $\beta_{m,k}$ is computed from the AP–UE geometry via the path-loss model~\cite{Venkatesh_CF_2022_Channel}. Using this, the approximate per-UE SINR is given by
\begin{align}
    \widetilde{\mathrm{SINR}}_k(\mathcal{T}) \approx 
\frac{ \beta^2_{\text{sum},k}(\mathcal{T})}{ \sum_{i \neq k} \beta^2_{\text{sum},i}(\mathcal{T}) + \tilde{\sigma}^2}.
\end{align}
Here $\tilde{\sigma}^2$ is the effective noise variance. The sum SE is given as $\widetilde{\mathrm{SE}}(\mathcal{T}) = \sum_{k=1}^K \log_2\!\big(1+\widetilde{\mathrm{SINR}}_k(\mathcal{T})\big)$. 
The next transmit AP is selected to maximize the incremental SE gain:
\begin{align}
t^\star \in \arg\max_{t\notin\mathcal{T}_i}  \big \{ \widetilde{\mathrm{SE}}(\mathcal{T}_i\cup\{t\}) - \widetilde{\mathrm{SE}}(\mathcal{T}_i)\big \}.
\end{align}

This process continues until  $|\mathcal{T}|=M_t$, after which the remaining APs are assigned as receivers. By construction, this scheme provides strong SE for the UEs. However, since the receiver AP set is only residually determined, the resulting geometry may be suboptimal for sensing, as the receivers can be spatially clustered.

\subsection{Sensing-centric Approach}
In the sensing-centric mode selection, the goal is to configure the receive APs such that they provide maximum geometric diversity across the coverage area. Unlike the communication-centric approach, which relies on AP–UE channel statistics, here the selection is determined solely by the receiving AP positions. In particular, a spatially separated set of receiver APs is expected to provide better geometric dilution of precision (GDOP)~\cite{Richard_GDOP}. 
This can be achieved  by a farthest-point sampling strategy, where receivers are iteratively chosen to maximize their separation from the already selected set. Given the set of receivers $\mathcal{R}_i$ at iteration $i$, we determine
\begin{align}
    r^\star \in \arg\max_{r\notin\mathcal{R}_i}\ \min_{r'\in\mathcal{R}_i} \left\|\mathbf{p}_r - \mathbf{p}_{r'}\right\|,\; \; 
\end{align}
then set $\mathcal{R}_{i+1}=\mathcal{R}_i\cup\{r^\star\}$ and repeat until $|\mathcal{R}|=M_r$. 
\section{Simulation Results}\label{sec: Simulation_results}\vspace{0pt}

In this section, we present numerical simulations to evaluate the proposed two-stage target estimation framework and to characterize the trade-off between communication and sensing under different AP mode selection strategies and power allocation weights $\rho_t$. 

\subsection{Scenario and Metrics}
We consider a scenario where $M=12$ APs, $S=2$  targets, and $K= \{4,10 \}$ UEs are randomly deployed within $1 \times 1 \text{ km}^2$ area.
The two dynamic targets are explicitly modeled, while static background clutter is assumed to be removed through calibration measurements collected in a target-free environment.
Each AP is equipped with a ULA having $N=8$ antennas, spaced at half-wavelength intervals $\lambda/2$. All the APs are assumed to be phase-synchronized, and the positions of the APs and UEs are perfectly known at the central processor \cite{Elisabetta_Matricardi_2025_DMIMO_OFDM,NON_COH_CHRISTOS}. The 2D positions of the  targets are unknown and can lie anywhere within the coverage area, and each target has an RCS of $0$ dBsm.
The system operates at a carrier frequency of $f_c = 3.5$~GHz with a narrow bandwidth of $W = 100$~kHz. The path loss for the communication channels follows the 3GPP Urban Microcell model ~\cite{Venkatesh_CF_2022_Channel}, while the sensing channels are modeled using the radar equation \cite{Zinat_24_Multi_Static_DMIMO}. The noise power is calculated as $\sigma^2 = k_B T_{\text{th}} W$, where $k_B$ is the Boltzmann constant, and $T_{\text{th}} = 290$~K is the thermal noise temperature. Each AP allocates communication power to UEs proportional to their channel strengths, with $\sum_k p_{tk} = \rho_t P_t$.

To evaluate the communication performance, we rely on the spectral efficiency derived in Section \ref{sec:commSE}, while the sensing performance is measured by the statistics of the error in the targets' position, the mean of which is in turn lower bounded by the position error bound\footnote{Details of the position-related Fisher information and PEB are omitted for space reasons.} (PEB)~\cite{Henk_Radio_Stripes_Journal_25}. Finally, the AP mode selection methods from Section \ref{sec:APselection} are evaluated in terms of the sensing coverage, defined as 
    $f_{\text{SC}}(\eta) = \text{Pr}(\text{PEB}(\mathbf{p}^s) \le \eta)$, 
where $\text{PEB}(\mathbf{p}^s)$ is the PEB of a target located in $\mathbf{p}^s$ and the probability considers the target location to be random within the deployment region.  


\subsection{Estimation Algorithm Performance Assessment }
\begin{figure}
    \centering  
    \includegraphics[width=0.8\linewidth]{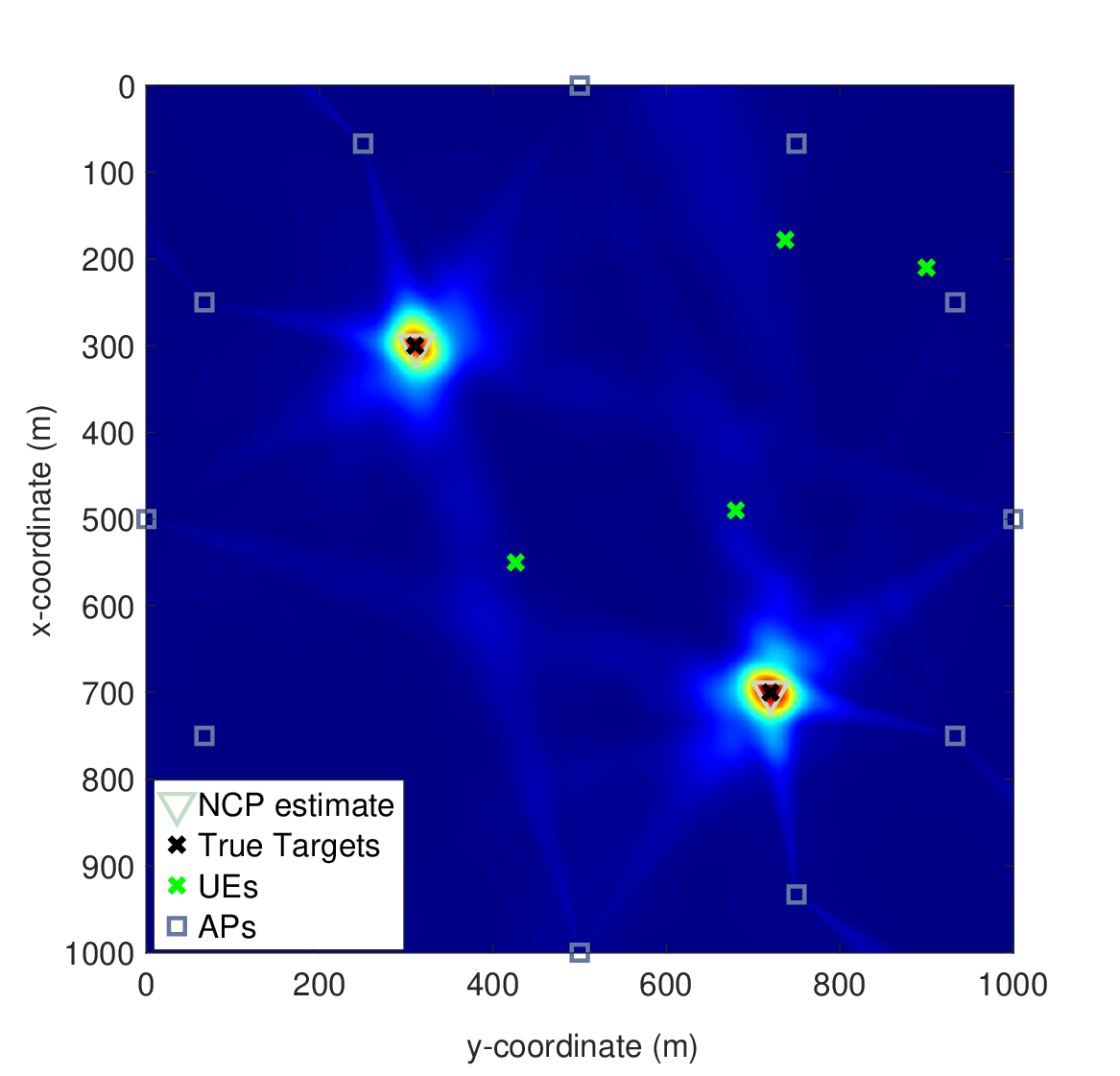}\vspace{-7pt}
        \caption{Non-coherent ML cost function with respect to the target’s $x$- and $y$-coordinates when the targets are well-separated. }
    \label{fig:cost_function_CP_NCP_far}    
\end{figure}

We begin by evaluating the proposed two-stage estimator, starting with the coarse detection step based on the non-coherent ML (NCP) estimator.
Fig.~\ref{fig:cost_function_CP_NCP_far} illustrates the NCP cost function in \eqref{eq-NCP_ml_2D} evaluated over the target’s $(x,y)$-coordinates where the two targets are well-separated. 
The NCP performs multi-target detection by scanning the spatial grid and evaluating the phase-insensitive 
cost function at each location. The resulting cost map exhibits distinct dips (negative peaks) corresponding to dominant reflections, which are identified using the CFAR test.  When the targets are well separated, as in Fig.~\ref{fig:cost_function_CP_NCP_far}, these dips appear at the true target locations and can be clearly distinguishable, enabling accurate detection for both targets. 
When the targets are closely spaced, the non-coherent estimation becomes limited by the spatial resolution. In this case, the reflected signals from nearby targets are highly correlated, causing their responses in the cost function to overlap. As a result, the non-coherent estimator cannot distinguish the individual targets and interprets them as a single  target. 



\begin{figure*}[h]
	\centering\vspace{+0.02in}
	\begin{subfigure}{.32\textwidth}
		\centering
		\includegraphics[width=\linewidth,height=0.92\linewidth]{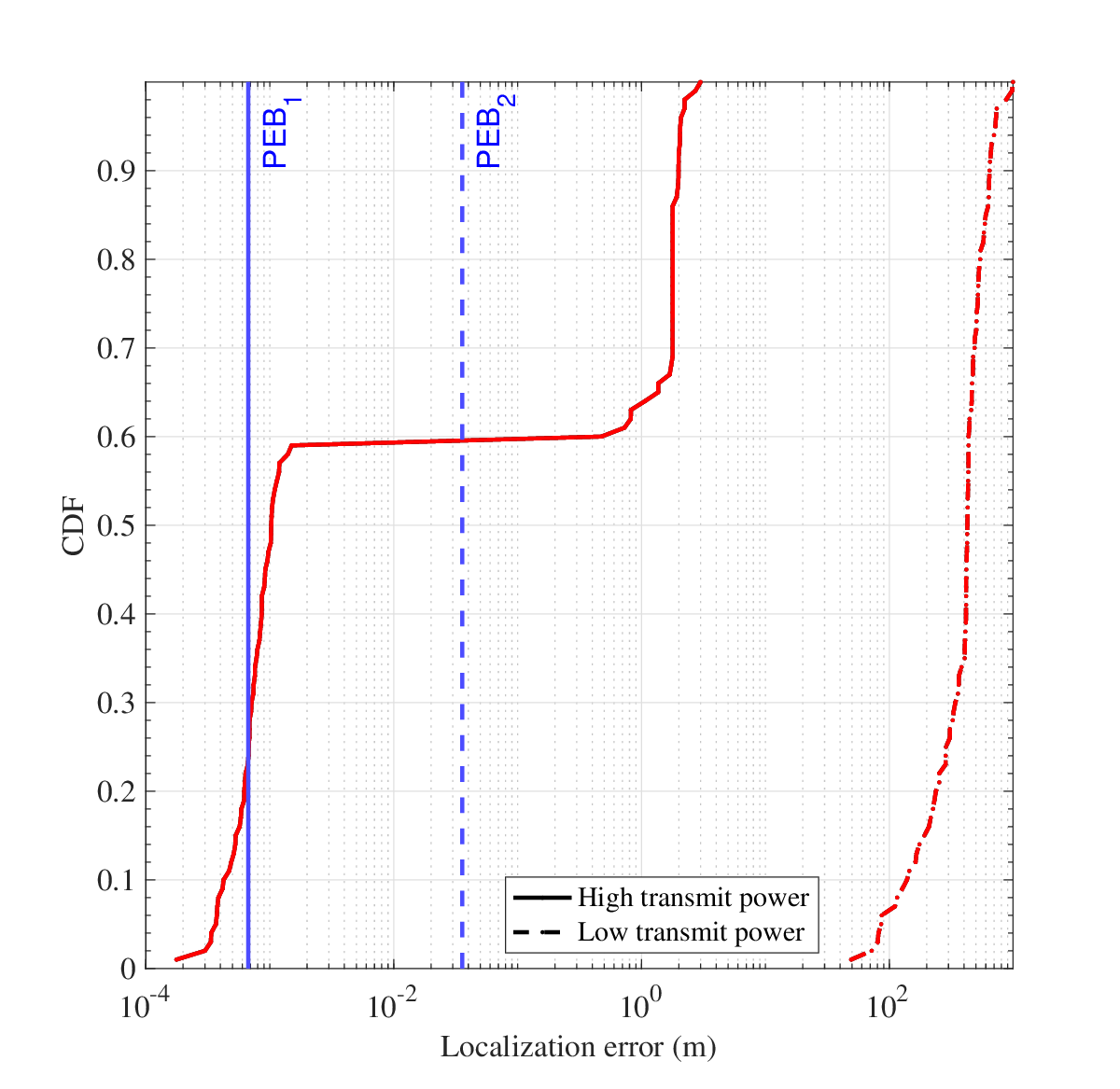}\vspace{-7pt}
		\caption{ \small} 
		\label{fig:CDF-PEB-RMSE}
	\end{subfigure}
	\begin{subfigure}{.32\textwidth}
		\centering
		\includegraphics[width=\linewidth,height=0.92\linewidth]{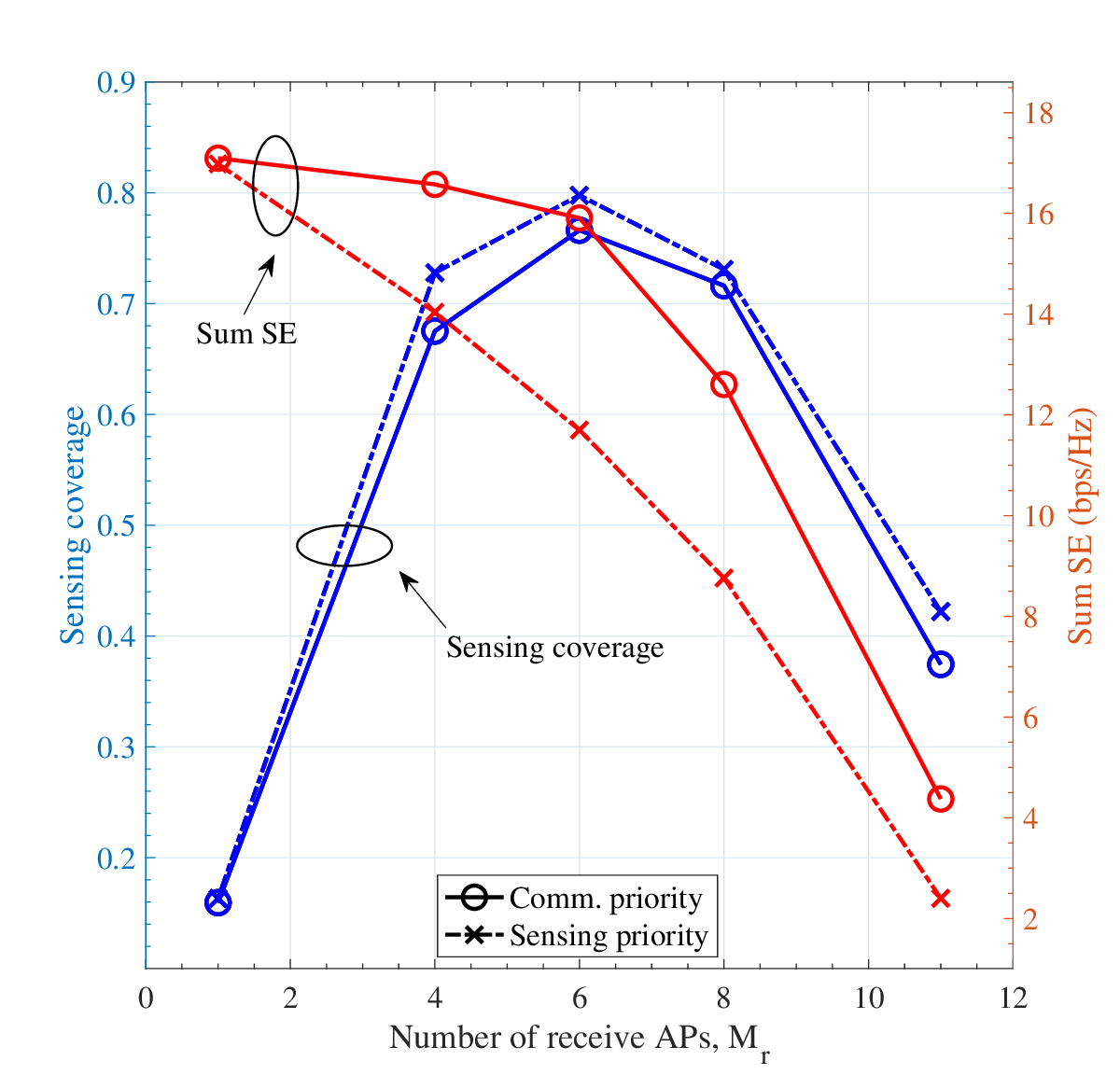}\vspace{-7pt}
		\caption{\small }
		\label{fig:SE-Coavrage_PEB_vary-APs} 
	\end{subfigure}
	\begin{subfigure}{.32\textwidth}
		\centering
		\includegraphics[width=\linewidth,height=0.92\linewidth]{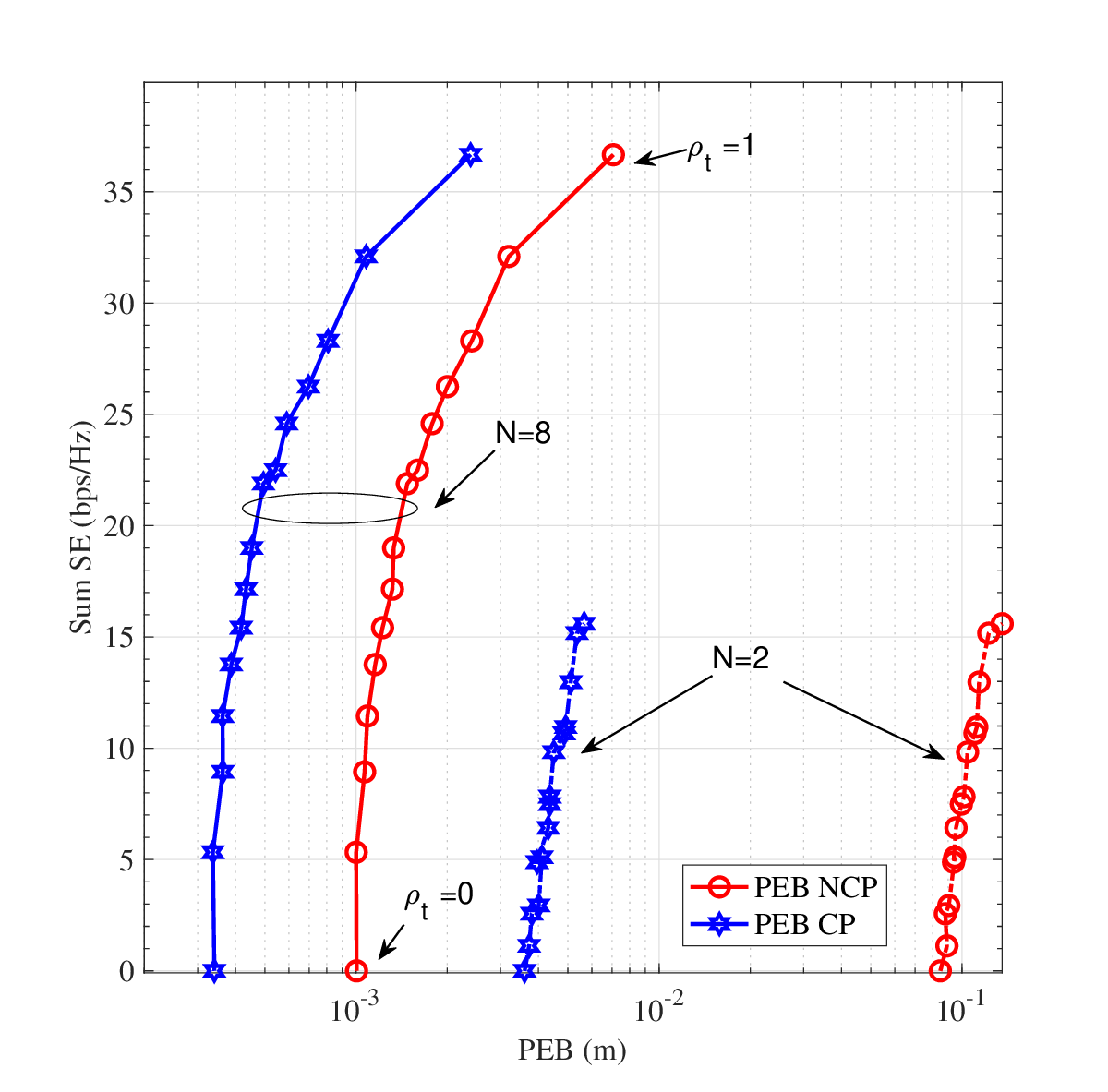}\vspace{-7pt}
		\caption{\small }
		\label{fig:SE-Rho_PEB}
	\end{subfigure}
	\vspace{-6pt}
	\caption{a) CDF of the multi-target positioning error obtained from coherent processing under high and low transmit-power settings $P_t = \{0, 40\}$~dBm; b) Sensing coverage $f_{\text{SC}}(\lambda/10)$ and sum SE versus the number of receive APs $M_r$ for a fixed $M=12$, showing the communication- and sensing-priority AP selection;  c) Sum SE versus PEB for different communication power-weight factor $\rho_t$ and antenna configurations. \vspace{-15pt}} 
	\label{fig:test}	
\end{figure*}

Using the coarse detections obtained from the non-coherent stage, Fig.~\ref{fig:CDF-PEB-RMSE} evaluates the coherent ML estimator's sensing performance by comparing the cumulative density function (CDF) of positioning error values for two targets in fixed locations from Fig.~\ref{fig:cost_function_CP_NCP_far}, randomizing the noise. 
The coherent estimator refines the target positions by exploiting the phase information across all synchronized APs to achieve high-accuracy sensing. 
At high transmit power,  the average positioning error approaches the PEB, indicating that the estimator efficiently utilizes the available phase information and operates close to the theoretical accuracy limit. In contrast, at low transmit power, noise limits the reliability of the non-coherent stage, resulting in inaccurate coarse estimates, while the spiky, non-convex ML cost surface further prevents the coherent refinement from converging to the global optimum.
Consequently, the coherent estimates deviate significantly from the true positions, and the average error value  remains well above the PEB.



\subsection{SE vs PEB Analysis}
We now, in Fig.~\ref{fig:SE-Coavrage_PEB_vary-APs}, evaluate the impact of the transmit/receiver split and compare communication-centric and sensing-centric AP mode-selection strategy on communication and sensing when the total APs are fixed at $M=12$. The APs are randomly deployed within the coverage area, and each AP transmits with $P_t = 20$~dBm. The communication power-weight factor is set to $\rho_c = 0.5$, with $N=8$ antennas, and $K=10$ UEs.
In Fig.~\ref{fig:SE-Coavrage_PEB_vary-APs}, the sensing coverage (left y-axis) quantifies the fraction of the area where the PEB is below the desired target, while the sum SE (right y-axis) measures the overall downlink communication throughput. For a given transmit and receive AP split $(M_t,M_r)$, the sensing coverage is computed as $f_{\text{SC}}(\eta = \lambda/10)$. The communication-centric strategy selects $M_t$ transmit APs to maximize the sum SE (solid red line), while the sensing-centric strategy selects the $M_r$ receive APs to improve the sensing geometry, yielding higher coverage PEB (dotted blue line).
When the number of receiving APs $M_r$ increases from $1$ to $6$, the sensing coverage initially improves since more APs operate in receive mode, providing a richer observation geometry for target estimation.  However, further increasing $M_r$ reduces the number of transmit APs ($M_t = M-M_r$), leading to lower transmit power and fewer phase-coherent illuminations, which degrades sensing performance.
In contrast, the sum SE decreases monotonically with $M_r$, as fewer APs contribute to data transmission.
Consequently, both sensing coverage and sum SE curves exhibit an opposite trend, highlighting \textit{the inherent communication–sensing trade-off in D-MIMO ISAC} systems. The intermediate region, around $M_r \approx 6$, achieves the best balance, offering high sensing coverage without severely compromising communication throughput.

Finally, in Fig.~\ref{fig:SE-Rho_PEB}, we assess the trade-off between communication and sensing performance by plotting the sum SE versus the PEB by varying the communication power weight factor $\rho_t$, with $M=12$ total APs and a fixed split of $M_t = M_r = 6$, and $P_t = 30$~dBm. We plot for  $ N=2$ and $N=8$  antennas per AP, and for both non-coherent and coherent PEB. 
For $N=2$, the array aperture is small, so the PEB remains in the centimeter range and changes slowly with $\rho_t$, whereas SE increases significantly with $\rho_t$ ($\approx 0 \rightarrow 15 $)~bps/Hz. There is also a large gap between the coherent and non-coherent PEB for $N=2$, because the coherent estimator exploits inter-AP phase information for positioning, while the non-coherent estimator relies only on geometric diversity through AoD/AoA information.  
For $N=8$, the larger aperture and array gain enhance both the SINR and the angular resolution, yielding higher SE and much smaller PEB values. Moreover, the PEB varies significantly, as we prioritize communication $\rho_t \uparrow$, moving from millimeter-level to centimeter-level accuracy. Importantly, around $\rho_t \approx 0.5$, the system still achieves millimeter-level sensing accuracy with reasonable SE, indicating a favorable operating point. Also, the coherent PEB is always smaller than the non-coherent PEB; this shows the benefit of exploiting phase information in the coherent scenario. Furthermore, the figure highlights the power-allocation trade-off: increasing $\rho_c$ improves SE but degrades PEB, with the trade-off being less severe for larger arrays ($N=8$). Hence, a system with more AP antennas can allocate a larger portion of transmit power to communication without significantly compromising sensing accuracy.






\vspace{-5pt}
\section{Conclusion}
This work investigated a D-MIMO ISAC framework that enables a trade-off between high communication SE and multi-target estimation using phase-coherent processing. The proposed two-stage ML-based estimator effectively combines non-coherent and coherent processing to achieve both robust target detection and sub-wavelength localization accuracy. In parallel, the AP mode-selection strategies provide a tunable trade-off between spectral efficiency and sensing coverage.   Numerical evaluations confirm that distributed phase coherent processing among APs achieves high-accuracy sensing.  We also showcased the SE–sensing trade-offs, achieving high SE with millimeter-level sensing accuracy under practical power-allocation constraints and moderate array size $N\!=\!8$. 
Future work will address imperfect synchronization, residual phase-offset calibration, dynamic multi-target tracking in cluttered environments, optimized AP mode selection, and wideband OFDM-based range estimation to advance the development of D-MIMO ISAC for~6G.

\balance
\vspace{-5pt}
\bibliographystyle{IEEEtran}
\bibliography{D_MIMO_ISAC_Venkatesh_References}
\end{document}